\documentclass{article}

	\usepackage[utf8]{inputenc}
	\usepackage[T1]{fontenc}
	\usepackage{graphicx}
	\usepackage{pifont,latexsym,ifthen,rotating,calc,textcase,booktabs,color}
	\usepackage{amsfonts,amssymb,amsbsy,amsmath,amsthm}
	\usepackage[errorshow]{tracefnt}
	\usepackage[top=3cm, bottom=3cm, left=3cm, right=3cm]{geometry}

	\title{Coupling of Finite-Element and Plane Waves Discontinuous Galerkin methods for time-harmonic problems}
	\author{M. Gaborit\textsuperscript{1}\textsuperscript{2}\textsuperscript{@},
	O. Dazel\textsuperscript{1}, P. G\"oransson\textsuperscript{2}
	G. Gabard\textsuperscript{1}}
	\date{1. LAUM, UMR CNRS 6613, Universit{\'e} du Maine, Le Mans, France\break
	2. MWL, KTH Royal Institute of Technology, Stockholm, Sweden\break
\textbf{@} gaborit@kth.se}

	\usepackage{textcomp}
	\newcommand\dd{~\mathrm{d}}
	\newcommand\mbf\mathbf
	\newcommand\dint{\int\displaylimits}
   	\newcommand\dx{\partial_x}
	\newcommand\dy{\partial_y}
	\newcommand\ff{\mathbf{f}}
	\newcommand\bb{\mathbf{b}}
	\newcommand\del{\delta}
	\newcommand\DDfull{\bb\left(\dx\ff,\dy\ff\right)}
	\newcommand\DD{\bb}
	
	\newcommand\vS{\mathbf{S}}
	\newcommand\vT{\mathbf{T}}
	
	\usepackage{amssymb}
	\usepackage{nicefrac}
    
    \newcommand\tb{\mbf}

    \newcommand{\Cin}{^{-}}
    \newcommand{\Cout}{^{+}}

	\usepackage{subcaption}

\begin{document}
	\maketitle
	\begin{abstract}
A coupling approach is presented to combine a wave-based method to the standard finite
		element method. This coupling methodology is presented here for the Helmholtz equation
		but it can be applied to a wide range of wave propagation problems. While wave-based
		methods can significantly reduce the computational cost, especially at high
		frequencies, their efficiency is hindered by the need to use small elements to resolve
		complex geometric features. This can be alleviated by using a standard Finite-Element
		Model close to the surfaces to model geometric details and create large, simply-shaped
		areas to model with a wave-based method. This strategy is formulated and validated in
		this paper for the wave-based discontinuous Galerkin method together with the standard
		finite element method. The coupling is formulated without using Lagrange multipliers
		and results demonstrate that the coupling is optimal in that the convergence rates of
		the individual methods are maintained.

	\end{abstract}

	\section{Introduction}

During the last decades simulations have become central in engineering and product design.
Driven by an increasing focus on reduced cost and shortened development time, a growing
number of industries rely on digital prototyping to guide their conceptual design process.
Initially used for research and earliest stages of concept definition, numerical models
and simulations gradually made their way towards the final stages of design and
validation. These trends result in an increasing need for fast, reliable and accurate
simulation techniques, with sufficient versatility to be used throughout the whole
development process.  The availability of large-scale computing resources (cloud-based
HPC, virtual clusters, etc... ) alleviated the problem for a while but as the models to be
simulated grow in size and detail, the need for computationally efficient methods remains
a priority.

These aspects are particularly relevant for wave propagation problems such as
acoustic models when using the Finite-Element Method (FEM). This tool, commonly acknowledged for its
robustness and adaptability, is suited for a wide range of problems as the use of
unstructured meshes allows handling complex geometrical details. On the other hand, this
method performs better in the low- and mid-frequency ranges but, in practice, is often
used for analysis at higher frequencies. However, it is well established that as the
frequency increases, the requirements in terms of elements per
wavelength~\cite{marburg_six_2002} induce a rapid increase of the size of the matrices
involved.  In addition, large gradients in the solution as well as geometrical details
often lead to a need for gradual mesh refinement in parts of the model.

It is beyond the scope of the present paper to review the research related to improve the
efficacy of dynamic FEM,  but a few examples that are related to the present work are e.g.
FE augmented with waves basis~\cite{babuska_partition_1997}, local
heuristics~\cite{babuska_generalized_2004}, high-order approximations in the shape
functions~\cite{beriot_analysis_2013,beriot_efficient_2016}, etc. In parallel, extensive
research has been conducted where alternative choices for interpolation and approximation
have been investigated, e.g. the Variational Theory of Complex
Rays~\cite{ladeveze_new_1996,riou_variational_2013} or waves such as the Wave Based
Method~\cite{desmet_wave_1998,deckers_wave_2014} which result in a reduced size of the
final linear system. The present work, focused on the Discontinuous Galerkin Method using
Plane Waves (PWDGM or DGM,~\cite{gabard_discontinuous_2007, gabard_comparison_2011}), is
part of this last category. This method uses a meshed domain and interpolates the fields
in each element using a basis of plane waves. Previous works demonstrated its ability to
compute solutions accurately over large, coarsely meshed domains with simple shapes. To
avoid numerical ill-conditioning due to linear dependence in the basis, the elements
should not become too small. Thus, the main drawback is related to complex geometries
where the mesh has to be refined to resolve geometrical details. In this case, as the
elements become smaller, the efficacy of the wave-based approach is lost. In order to
alleviate these limitations while still benefiting from the small-sized linear system,
there  is a growing interest in developing hybrid schemes where these advanced methods can
be coupled with standard FEM. Such attempts have been published for the
WBM~\cite{van_hal_hybrid_2004,lee_direct_2016} and the VTCR~\cite{ladeveze_trefftz_2014},
both for a coupling with standard FEM.

In the present paper, an original coupling technique for PWDGM and FEM is proposed. The
objective is to use both methods in an efficient way, i.e. using FEM to model geometrical
details and create large, simply-shaped areas to be accounted for using PWDGM. The key
towards efficacy in such a coupling technique is the way the coupling conditions between
the two domains are formulated. As an example, the authors have investigated an approach
based on the use of the derivatives of the FE shape functions to match DGM approach,
inducing the loss of one order of magnitude on the convergence
rate~\cite{dazel_couplage_2015}. The approach proposed in this paper avoids the use of
Lagrange multipliers and the associated increase in the number of equations as well as the
derivation of shape function. It is instead focused on mimicking the classical approaches
traditionally used within each of the method to handle boundary conditions. A prominent
feature of the proposed coupling strategy is that it handles incompatible meshes without
any noticeable loss of numerical precision. Thus, no particular care has to be exercised
to ensure the nodes of the two meshes match at the interface. Finally, the results show
that this coupling strategy does not induce additional numerical error. On the contrary,
the results suggest that the maximum error level is controlled by the highest error of the
two coupled methods.

The paper begins with an overview of both methods given in section~\ref{sec:basics},
followed by a theoretical description of the proposed coupling procedure in
section~\ref{sec:coupling}. While the proposed approach is general and may be applied to
couple different physical media on both sides of the interface, in this paper the
procedure is demonstrated on a simple fluid-fluid case and applied to academic examples as
well as a more complex and challenging test case.

Throughout the paper a harmonic time dependence is assumed with an implicit convention
$e^{j\omega t}$. The fields are represented by their respective complex amplitude.

\section{Basics}
\label{sec:basics}

\begin{figure} 
	\centering
	\includegraphics{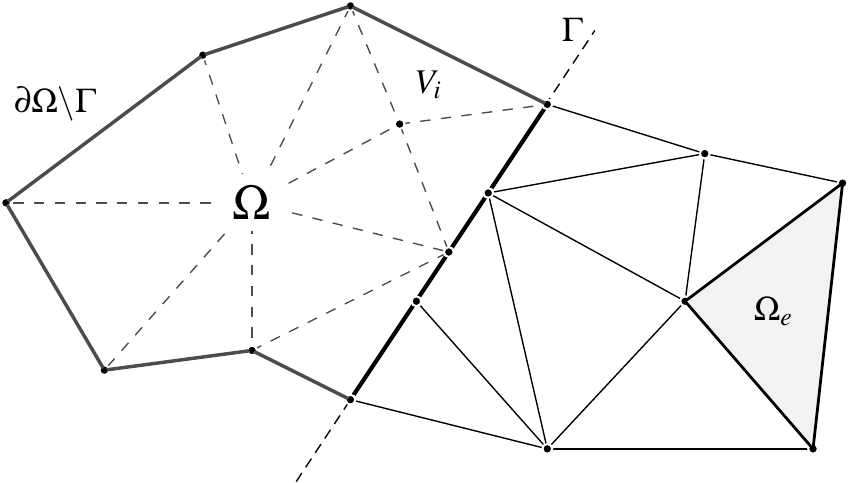}
	\caption{The two domains 1 and 2 are modelled with different methods, coupled through
	the interface $\Gamma$. The meshes from both sides are not necessarily consistent.}
	\label{fig:problem_position}
\end{figure}

The system considered (see figure~\ref{fig:problem_position}) is composed of two media
(denoted 1 and 2) sharing an interface $\Gamma$. For every point along $\Gamma$, relations
between the physical fields in the surrounding media may be established. As some of the
fields may vanish at the interface, as with essential boundary conditions, the term
\textit{interface relations} will be preferred over continuity conditions. These relations
may be expressed as a matrix equation in the following general form:

\begin{equation}
	\begin{bmatrix}
    \mbf{C}_1(x)
    \end{bmatrix}\mbf{X}_1(x)=\begin{bmatrix}
    \mbf{C}_2(x)
    \end{bmatrix}\mbf{X}_2(x), \quad\forall\, x\in\Gamma
    	\label{eq:coupling:localgeneral}
\end{equation}
where the vectors $\mbf{X}_i(x)$ represent  the complex amplitudes of the physical fields
involved in the interface relation for each medium $i$ and in each point $x$ of $\Gamma$.
The proposed coupling approach handles both different physical models and numerical
schemes, hence $\mbf{X}_1(x)$ and $\mbf{X}_2(x)$ used to represent the media's dynamic
state may be different.  The problem is well-posed if the interface relations are known
and sufficient to uniquely determine the solution. From now on, the spatial dependence $x$
in the relation~\eqref{eq:coupling:localgeneral} will be omitted for conciseness.

In order to facilitate the discussion, the weak forms for FEM and DGM are presented. As
the DGM is more recent, it will be presented in greater detail. Since the proposed
coupling procedure is independent of the chosen discretization and interpolation schemes,
these are not discussed here. The objective of this section is to introduce the boundary
terms and composition of the $\mbf{X}_i(x)$ vector involved for each scheme.

\subsection{Finite Element Method}
	\label{sec:basics:fem}

In this section, the FE formulation for a second order partial differential equation is presented. The problem is set on a domain $\Omega$ whose boundary $\partial\Omega$ is
assumed to be sufficiently regular. The associated weak form reads:

\begin{equation}
	\dint_\Omega a(\ff,\del\ff)\dd\Omega +
	\dint_\Omega L(\del\ff) \dd\Omega +
	\dint_{\partial\Omega} \mbf{b}(\dx\ff, \dy\ff)\del\ff \dd\Gamma = 0
	~,\quad \forall\del\ff\in \mathcal{V}
	\label{eq:fem:wf}
\end{equation}
where $\ff$ denotes the vector of primary variables of the weak form and $\del\ff$ the
associated test fields. The $\ff$ vector belongs to a Hilbert space $\mathcal{V}$ and
$\del\ff$ to its dual. The variables in $\ff$ are
discretized over the mesh. The bilinear function $a$ models the reaction of the media in
the volume and depends on the fields and their spatial derivatives. When the weak form is
discretized and written as a linear system, $a$ provides the system matrix.
The inner volume forces acting on the medium are given by $L$.

The interface relations are introduced through the function $\bb$, representing the
interface relations through combinations of the so-called secondary variables. Together
with the primary variables $\bb$ forms the $\mbf{X}$ vector used in~\eqref{eq:coupling:localgeneral}.

As an example, a common choice for a fluid medium uses the scalar acoustic pressure $p$ as
primary variable and the normal velocity as secondary. Hence:
\begin{equation}
\tb{X}_1=\begin{Bmatrix}
\DDfull\\
\ff
\end{Bmatrix}=\begin{Bmatrix}
	\left(\dfrac{\nabla p\cdot\mbf{n}}{j\rho\omega}\right)\\
	p
\end{Bmatrix}
	\label{eq:fem:bo_fluid}
\end{equation}
with $\rho$ denoting the medium's density and $\nabla\bullet\cdot\mbf{n}$ the normal derivative.

	\subsection{Discontinuous Galerkin Method}
	\label{sec:basics:dgm}

To formulate the DGM a common first step is to express the dynamic behaviour as a system
of first order differential equations and then proceed towards a spatially
discretized weak form~\cite{gabard_discontinuous_2007}. The problem is then modelled by:

	\begin{equation}
		j\omega\vS
		+ [\mbf{A}]\dx\vS
		+ [\mbf{B}]\dy\vS
		= \mbf{0}
		\label{eq:dgm:governing}
	\end{equation}
where $\vS$, usually called state vector, is a vector whose components are combinations of
the physical fields from $\ff$ and their space derivatives. Different expressions for
$\vS$, $[\mbf{A}]$ and $[\mbf{B}]$ can be
used~\cite{gabard_discontinuous_2007,gabard_discontinuous_2015}.  As an example, for a
fluid medium, the following state vector will be used:
\begin{equation}
	\vS=\Big\{v_x,~v_y,~p\Big\}^T
\end{equation}

A key aspect of the DGM is that fields are approximated as continuous-per-element and
only the flux continuity is enforced across the elements. To use this approach, the domain
has to be meshed and the weak form to be decomposed over it.
A set of elementary sub-domains $\Omega_e$ (with $e=1,\ldots,N$) is then introduced and the
weak form distributed over the union of all $\Omega_e$.

\begin{equation}
	 \sum_{e=1}^{N} \dint_{\Omega_e} \vT_e^T\Big(
	j\omega
	+ [\mbf{A}_e]\dx
	+ [\mbf{B}_e]\dy
	\Big)\vS_e ~\dd\Omega
	= 0,\quad \forall\vT_e\in\mathcal{S}_e
	\label{eq:dgm:elementwise}
\end{equation}
where $\vS_e$ and $\vT_e$ correspond respectively to the state variables and test
functions over the element $\Omega_e$. The vector of unknowns, $\vS_e$, is identical to
the state vector previously introduced. The vector of test functions, on the other hand,
belong to an Hilbert space $\mathcal{S}_e$, constructed as the product of Sobolev spaces
in which lie the different physical fields descriptors.

In order to separate out boundary and interior terms, the following identity may be used:

\begin{equation}
	\dint_{\Omega_e}
		\nabla \cdot
		\begin{Bmatrix}
			\vT_e^T [\mbf{A}_e] \vS_e\vspace{5pt}\\
			\vT_e^T [\mbf{B}_e] \vS_e
		\end{Bmatrix}
	\dd\Omega
	=
	\dint_{\partial\Omega_e}
		\begin{Bmatrix}
			\vT_e^T [\mbf{A}_e] \vS_e\vspace{5pt}\\
			\vT_e^T [\mbf{B}_e] \vS_e
		\end{Bmatrix}
		\cdot \mbf{n}
	\dd\Gamma
    \label{eq:dgm:stokes}
\end{equation}
with the vector $\mbf{n}$ is the normal to $\partial\Omega_e$ pointing outwards $\Omega_e$.
When evaluating~\eqref{eq:dgm:stokes}, the derivatives related to $\vS_e$ may be eliminated using~\eqref{eq:dgm:elementwise}, leading to:

\begin{equation}
	\sum_{e=1}^{N} \dint_{\Omega_e} \Big(
		j\omega\vT_e
		- [\mbf{A}_e]^T\dx\vT_e
		- [\mbf{B}_e]^T\dy\vT_e
	\Big)^T\vS_e\dd\Omega
	+ \sum_{e=1}^{N} \dint_{\partial\Omega_e} \vT_e^T[\mbf{F}_e]\vS_e\dd\Gamma
	= 0,\quad \forall \vT_e\in \mathcal{S}_e
	\label{eq:dgm:postgreen}
\end{equation}
with $[\mbf{F}_e] = [\mbf{A}_e]n_x + [\mbf{B}_e]n_y$, being the \textit{flux matrix} associated with direction $\mbf{n} = \{n_x,~n_y\}^T$.

The final step towards the derivation of the DGM equation is to eliminate the volume
integral in~\eqref{eq:dgm:postgreen}. This is achieved by choosing the test functions
among the solutions to the adjoint of~\eqref{eq:dgm:governing}. Thus,
taking $\vT_e$ in a subspace $\mathcal{T}_e \subset \mathcal{S}_e$ such as:

\begin{equation}
	\mathcal{T}_e = \Big\{
		\vT_e \in \mathcal{S}_e \, \Big| \,
		j\omega\vT_e
		- [\mbf{A}_e]^T\dx\vT_e
		- [\mbf{B}_e]^T\dy\vT_e = \mbf{0}
	\Big\}
	\label{eq:dgm:adjoint}
\end{equation}
allows for the volume integrals of~\eqref{eq:dgm:postgreen} to be cancelled,
leaving only a sum of integrals over the element boundaries:

\begin{equation}
	\sum_{e=1}^{N} \dint_{\partial\Omega_{e}} \vT_{e}^T[\mbf{F}_{e}]\vS_{e}\dd\Gamma
	= 0~,\qquad \forall \vT_{e}\in \mathcal{T}_{e}
	\label{eq:dgm:postchoice}
\end{equation}

Equation~\eqref{eq:dgm:postchoice} addresses both internal interfaces of the DGM
domain and its external boundaries. The part of this equation treating with internal
interfaces might be rewritten as a sum on those by introducing $\Gamma_{ee'}$, interface
between the element $e$ and $e'$:

\begin{equation}
	\sum_{e, e'<e} \int_{\Gamma_{ee'}} \vT_e^T[\mbf{F}_e]\vS_e + \vT_{e'}^T[\mbf{F}_{e'}]\vS_{e'}\dd\Gamma
\end{equation}

A basic requirement for the present formulation to be conservative is that the normal
fluxes from both sides of $\Gamma_{ee'}$ equal at the interface:

\begin{equation}
	[\mbf{F}_e]\vS_e+[\mbf{F}_{e'}]\vS_{e'}=0
\end{equation}

No condition on the $\vS_{e}$ and $\vS_{e'}$ vectors has been set to enforce this
condition. To this end, a numerical flux $\mbf{f}_{ee'}(\vS_e, \vS_{e'})$ might be
constructed as proposed in the literature,
e.g.~\cite{gabard_discontinuous_2007,gabard_discontinuous_2015}. This flux is then split
into incoming and outgoing parts and the former is used to express the latter.

On the other hand, a set of external boundaries comes to complete the formulation and
reads:

\begin{equation}
	\sum_{e} \int_{\partial\Omega_e\cap\partial\Omega}\vT_e^T[\mbf{F}_e]\vS_e\dd\Gamma
	\label{eq:dgm:external_boundaries}
\end{equation}

The coupling procedure aims at proposing a new form for the flux
$[\mbf{F}_e]\vS_e$ between the DGM and the FEM domains to substitute in the relevant terms
of the summation~\eqref{eq:dgm:external_boundaries}.

	\section{Coupling}
	\label{sec:coupling}

To establish the coupling between the weak forms, the sum of the weak forms
~\eqref{eq:fem:wf} and ~\eqref{eq:dgm:postchoice} leads to:

\begin{align}
	\dint_\Omega a(\ff, \mbf{\delta f}) \dd\Omega+
	\dint_\Omega L(\mbf{\delta f}) \dd\Omega +
	\sum_{e=1}^{N}
		\dint_{\partial\Omega_{e}\backslash \Gamma}\vT_{e}^T[\mbf{F}_{e}]\vS_{e}\dd\Gamma
		+ \dint_{\partial\Omega\backslash\Gamma} \mbf{b}\left(\dx\ff,\dy\ff\right)\mbf{\delta f} \dd\Gamma
	\notag\\
	+ \dint_{\Gamma} \mbf{b}\left(\dx\ff,\dy\ff\right)\mbf{\delta f} \dd\Gamma
	+ \sum_{e=1}^{N'} \dint_{\Gamma}\vT_{e}^T[\mbf{F}_{e}]\vS_{e}\dd\Gamma = 0
	~,\quad \forall\mbf{\delta f}\in \mathcal{V}, \vT_{e}\in\mathcal{T}_e
	\label{eq:coupling:full}
\end{align}
where the integrals corresponding to the interface between the two sub-domains are
evaluated along $\Gamma$ and the rest of the section is focused on these terms.

\begin{figure} 
	\centering
	\includegraphics{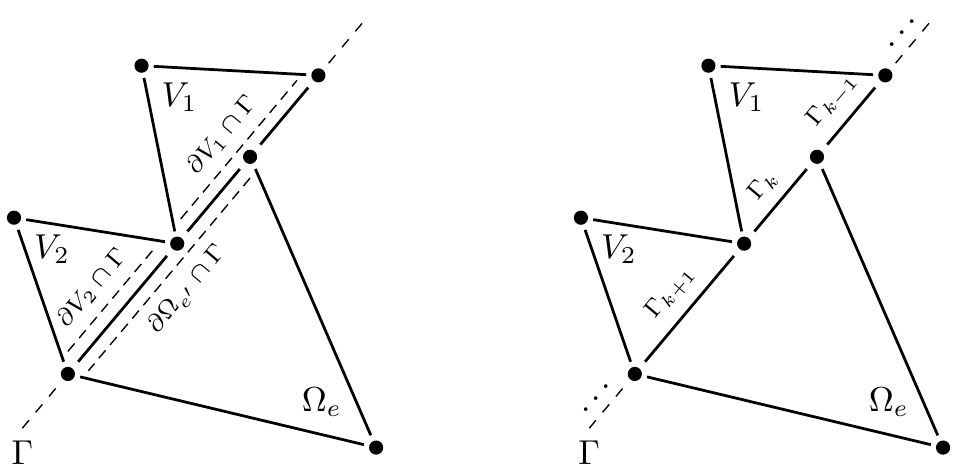}
	\caption{Renumbering the different segments of the interface $\Gamma$ with $\Gamma_k$ and $k$
	from $1$ to $N_k$. This new scheme verifies $\bigcup_{k_=1}^{N_k}\Gamma_k
	= \Gamma$.}
	\label{fig:ifacerenumbering}
\end{figure}

Both domains require some form of meshing to be applied and the interface will thus be
represented twice, however these meshes need not to be compatible. The union of these two
discretizations splits $\Gamma$ into several segments.  To express the coupling terms, an
efficient way of addressing each segment of $\Gamma$ is needed.  As shown in
figure~\ref{fig:ifacerenumbering}, each of the $N_k$ segments of $\Gamma$ connecting two
consecutive nodes is denoted $\Gamma_k$ with $1 \leq k \leq N_k$. This set is built in
such a way that it covers the whole $\Gamma$ interface:

\begin{equation*}
	\Gamma=\bigcup_{k=1}^{N_k}~\Gamma_k
\end{equation*}

All segments have their normal oriented outwards from the DG domain.  For each them, it is
straightforward to identify the surrounding FEM and DGM elements, hence the element
indices are omitted in the following for improved readability.  The second line of
(\ref{eq:coupling:full}) can then be rewritten as:

\begin{equation}
	I_C =
		\sum_{k=1}^{N_k} \bigg(
			\dint_{\Gamma_k}\mbf{b}\left(\dx\ff,\dy\ff \right)\mbf{\delta f}\dd\Gamma +
			\dint_{\Gamma_k}\vT^T[\mbf{F}]\vS\dd\Gamma
		\bigg)
	\label{eq:coupling:Ic}
\end{equation}

The key idea behind the proposed method is to derive explicit, local expressions for $\bb$
and the flux $[\mbf{F}]\vS$ as functions of the primary variables and the state vector
$\vS$. The starting point to derive such relations is obtained from a substitution of
(\ref{eq:fem:bo_fluid}) into  (\ref{eq:coupling:localgeneral}) thus expressing physical
continuity conditions between the DGM state vector and the FEM variables ($\ff$ and
$\DD$):

\begin{equation}
	[\mbf{C}_1]\begin{Bmatrix}
		\DDfull\\
		\ff
	\end{Bmatrix}
	=
	[\mbf{C}_2] \vS
	\label{eq:coupling:continuity}
\end{equation}
where $[\mbf{C}_1]$ and $[\mbf{C}_2]$ are matrices combining the different quantities to
represents valid interface relations.

The flux entering the DGM sub-domain is separated from the one leaving it by introducing
the characteristics of the differential operator.

Given that the eigenspace of $[\mbf{F}]$ is identical to the space of characteristics, the
decomposition into outgoing and incoming flux may be obtained by diagonalizing the flux
matrix: $[\mbf{F}]=[\mbf{P}][\mbf{\Lambda}][\mbf{Q}]$.  In this relation $[\mbf{P}]$
represents the matrix of eigenvectors and $[\mbf{Q}]$ its inverse. The eigenvalues (on the
diagonal of $[\mbf{\Lambda}]$) can then be separated in two sets.  The positive (resp.
strictly negative) ones, associated with characteristics going out of the element (resp.
entering in to) are denoted with a $+$ (resp.  $-$). It is important to stress that the
zero-valued characteristics are non propagative and that they will have no role in
rewriting the interface relations. Despite this, for consistency, they are kept and
grouped together with the positive ones. This leads to the following partitioning of the
matrices $[\mbf{P}]$ and $[\mbf{Q}]$:

\begin{equation}
  [\mbf{P}] = \left[\, [\mbf{P}\Cin] \,|\, [\mbf{P}\Cout] \,\right]
  ,\qquad
  [\mbf{Q}] = \begin{bmatrix}
  	[\mbf{Q}\Cin]\\
  	[\mbf{Q}\Cout]
  \end{bmatrix},
\end{equation}

with the following decomposition for the State vector $\vS_e:$

\begin{equation}
	\vS =[\mbf{P}\Cin]\vS\Cin+[\mbf{P}\Cout]\vS\Cout
	,\qquad
    \vS\Cin =[\mbf{Q}\Cin]\vS
    ,\qquad
    \vS\Cout =[\mbf{Q}\Cout]\vS
	\label{eq:coupling:S:decomposition}
\end{equation}
where $\vS\Cin$ and $\vS\Cout$ are the generalized coordinates of $\vS$ in the space of
characteristics, forming an intermediate step towards determining flux compliant with the
boundary conditions. Introducing $\vS\Cin$ and $\vS\Cout$ into
(\ref{eq:coupling:continuity}), and partitioning $[\mbf{C}_1]$ into sub-matrices related
to primary or secondary variables:

\begin{equation}
	\Big[~
		[\mbf{C}^{\DD}_1]
	~\Big|~
		[\mbf{C}^{f}_1]
	~\Big]
	\begin{Bmatrix}
		\DDfull\\
		\ff
	\end{Bmatrix}
	=
	\Big[~
		[\mbf{C}_2][\mbf{P}\Cin]
	~\Big|~
		[\mbf{C}_2][\mbf{P}\Cout]
	~\Big]
	\begin{Bmatrix}
		\vS\Cin\\
		\vS\Cout
	\end{Bmatrix}
\end{equation}

Re-arranging this equation in order to gather terms related to $\DD$ and $\vS\Cin$ on one
side and $\ff$ and $\vS\Cout$ on the other side:

\begin{equation}
	\Big[~
		[\mbf{C}^{\DD}_1]
	~\Big|~
		-[\mbf{C}_2][\mbf{P}\Cin]
	~\Big]
	\begin{Bmatrix}
		\DDfull\\
		\vS\Cin
	\end{Bmatrix}
	=
	\Big[~
		-[\mbf{C}^{f}_1]
	~\Big|~
		[\mbf{C}_2][\mbf{P}\Cout]
	~\Big]
	\begin{Bmatrix}
		\ff\\
		\vS\Cout
	\end{Bmatrix}
	\label{eq:coupling:reordered}
\end{equation}

Finally, after inverting the matrix at the left hand side:

\begin{equation}
	\begin{Bmatrix}
		\DDfull\\
		\vS\Cin
	\end{Bmatrix}
	=
	\Big[~
		[\mbf{C}^{\DD}_1]
	~\Big|~
		-[\mbf{C}_2][\mbf{P}\Cin]
	~\Big]^{-1}
	\Big[~
		-[\mbf{C}^{f}_1]
	~\Big|~
		[\mbf{C}_2][\mbf{P}\Cout]
	~\Big]
	\begin{Bmatrix}
		\ff\\
		\vS\Cout
	\end{Bmatrix}
	\label{eq:coupling:core}
\end{equation}

For convenience, the matrix linking the vectors of variables is called \textit{reflection
matrix} and denoted $[\mbf{R}]$. It is partitioned into four blocks numbered $[\mbf{R}_{ij}]$
which all link two components of the left and right hand vectors
of~\eqref{eq:coupling:core}:

\begin{equation}
	[\mbf{R}]
	=
	\Big[~
		[\mbf{C}^{\DD}_1]
	~\Big|~
		-[\mbf{C}_2][\mbf{P}\Cin]
	~\Big]^{-1}
	\Big[~
		-[\mbf{C}^{f}_1]
	~\Big|~
		[\mbf{C}_2][\mbf{P}\Cout]
	~\Big]	=
	\Bigg[
		\begin{matrix}
			[\mbf{R}_{11}] & [\mbf{R}_{12}]\\
			[\mbf{R}_{21}] & [\mbf{R}_{22}]
		\end{matrix}
	\Bigg]
	\label{eq:coupling:R}
\end{equation}

To assess the existence of $[\mbf{R}]$, one can check that the dimensions of
$[\mbf{P}\Cin]$, $[\mbf{C}_2]$ and $[\mbf{C}^b_1]$ make the second matrix
of~\eqref{eq:coupling:R} square. As the well-posedness is well-posed this matrix has to be
full rank, thus invertible.

\bigskip

Using~\eqref{eq:coupling:S:decomposition}, a new expression for $\DDfull$ taking into
account the interface conditions is formed. This expression shows explicit dependence on
$\ff$ and $\vS$, and thus are the objectives set forth in this paper accomplished.

\begin{equation}
	\DDfull=[\mbf{R}_{11}] \ff + [\mbf{R}_{12}][\mbf{Q}\Cout] \vS
	\label{eq:coupling:intfem}
\end{equation}

Based on the results obtained above, it is possible to modify the DGM flux $[\mbf{F}]\mbf{S}$ following the same decomposition approach. 
Introducing ~\eqref{eq:coupling:S:decomposition} to separate incoming and outgoing characteristics, as well as ~\eqref{eq:coupling:core} and \eqref{eq:coupling:R}:

\begin{align}
	[\mbf{F}]\vS = & [\mbf{F}]\Big([\mbf{P}\Cin]\vS\Cin + [\mbf{P}\Cout]\vS\Cout\Big)\notag\\
			& [\mbf{F}][\mbf{P}\Cin][\mbf{R}_{21}]\ff +
			[\mbf{F}]\Big([\mbf{P}\Cin][\mbf{R}_{22}] + [\mbf{P}\Cout]\Big)[\mbf{Q}\Cout]\vS
\label{eq:coupling:intdgm}
\end{align}

Finally, combining~\eqref{eq:coupling:intfem} and~\eqref{eq:coupling:intdgm} for all the segments $\Gamma_k$ a new form for the coupling operator across $\Gamma$ emerges:

\begin{align}
	I_C = \sum_{k=1}^{N_k} & \Bigg(
		\dint_{\Gamma_k} \mbf{\delta f} [\mbf{R}_{11}] \ff \dd\Gamma
		+ \dint_{\Gamma_k} \mbf{\delta f} [\mbf{R}_{12}][\mbf{Q}\Cout] \vS \dd\Gamma\notag\\
		& + \dint_{\Gamma_k} \vT^T [\mbf{F}][\mbf{P}\Cin][\mbf{R}_{21}]\ff \dd\Gamma
		+ \dint_{\Gamma_k} \vT^T [\mbf{F}]\Big(
					[\mbf{P}\Cin][\mbf{R}_{22}] +
        	[\mbf{P}\Cout]
        \Big) [\mbf{Q}\Cout]\vS\dd\Gamma
	\Bigg)
	\label{eq:coupling:final}
\end{align}

This new operator accounts for the interface conditions~\eqref{eq:coupling:continuity} using only
variables already present in the weak forms, enforcing the transmission of quantities
across $\Gamma$ without introducing additional variables such as e.g. Lagrange multipliers.
The coupling of the sub-domains is mainly provided through the second and third
terms of~\eqref{eq:coupling:final} with boundary reactions in the form of the first and fourth terms.

	\subsection{Application to a fluid-fluid case}

In order to give a better understanding of the method, this section focuses on an
analytical acoustic example. The reflection matrix is deduced for fluid media and the
method's consistency is demonstrated on a simple example.

The example involve two fluid media whose physical properties are $(\rho_i, c_i),~i\in\{1,2\}$
with $\rho_i$ the density of medium $i$ and $c_i$ its associated sound speed. The
characteristic impedance for the medium $i$ is expressed as $Z_i = \rho_ic_i$.

The first domain (subscripted $1$) is modelled by FEM, therefore the field of
representation used is the pressure $p$ and the boundary operator is described using the
normal velocity as proposed in equation~\eqref{eq:fem:bo_fluid}:

\begin{equation}
	\bb(\dx p_1, \dy p_1)  = -\frac{\nabla p_1\cdot\mbf{n}}{j\omega\rho_1} = \mbf{v_1}\cdot\mbf{n}
\end{equation}

In the other fluid medium (subscripted $2$), DGM is used and the state vector chosen is: $\vS_2 =
\{v_{x2},~v_{y2},~p_2\}^T$. One can then deduce the two matrices $[\mbf{A}_2]$,
$[\mbf{B}_2]$ representing the conservation equations:

\begin{equation}
	[\mbf{A}_2] = \begin{bmatrix}
		0 & 0 & \nicefrac{1}{\rho_2}\\
		0 & 0 & 0\\
		\rho_2c_2^2 & 0 & 0
	\end{bmatrix}~,\qquad
	[\mbf{B}_2] = \begin{bmatrix}
		0 & 0 & 0\\
		0 & 0 & \nicefrac{1}{\rho_2}\\
		0 & \rho_2c_2^2 & 0
	\end{bmatrix}
	\label{eq:ex:matrices}
\end{equation}

In order to evaluate the reflection matrix $[\mbf{R}]$, one needs to decompose the state
vector in terms of characteristics. The flux matrix $[\mbf{F}_2] = [\mbf{A}_{2}]n_x +
[\mbf{B}_{2}]n_y$ is then computed and diagonalized to extract the characteristics. The
two eigenvectors $[\mbf{P}_2\Cin]$ and $[\mbf{P}_2\Cout]$ associated with the phase speeds
$c_2$ and $-c_2$ are then retrieved:

\begin{gather}
	[\mbf{P}_2\Cin] = \begin{Bmatrix}
		-n_x\\ -n_y\\ Z_2
	\end{Bmatrix}~,\,
	[\mbf{P}_2\Cout] = \begin{bmatrix}
		{n_x} & {-n_y}\\ {n_y} & {n_x}\\ {Z_2} &{0}
	\end{bmatrix}~,\notag\\
	[\mbf{Q}_2\Cin] = \begin{bmatrix}
		-\dfrac{n_x}{2} & -\dfrac{n_y}{2}& \dfrac{1}{2Z_2}
	\end{bmatrix},\,
	[\mbf{Q}_2\Cout] =  \begin{bmatrix}
		\dfrac{n_x}{2} & \dfrac{n_y}{2}& \dfrac{1}{2Z_2}\\
        {-n_y} &n_x & 0
	\end{bmatrix}
	\label{eq:ex:flux_splitting}
\end{gather}

The reflection matrix consists mainly in a particular rewriting of continuity conditions.
For a fluid/fluid interface, interface relations enforce continuity of pressure and normal
veclocity which leads to the following form of~\eqref{eq:coupling:continuity}:

\begin{equation}
	\begin{Bmatrix}
		\bb(\dx p_1, \dy p_1)\\
		p_1
	\end{Bmatrix}
	=
	\begin{bmatrix}
		n_x& n_y& 0\\
		0& 0& 1
	\end{bmatrix}
	\begin{Bmatrix}
		v_{x2}\\
		v_{y2}\\
		p_2
	\end{Bmatrix}
\end{equation}

As proposed in the current work, $\vS_2$ is replaced by its representation in the space of
characteristics see ~\eqref{eq:coupling:S:decomposition}.
Reordering the terms leads to:

\begin{align}
	\begin{bmatrix}
		1 & 1\\
		0 & -Z_2
	\end{bmatrix}
	\begin{Bmatrix}
		\bb(\dx p_1, \dy p_1)\\
		\vS_2\Cin
	\end{Bmatrix}
	=
	\begin{bmatrix}
		0& 1 & 0\\
		-1& Z_2&0
	\end{bmatrix}
	\begin{Bmatrix}
		p_1\\
           \vS_2\Cout
	\end{Bmatrix}
	\label{eq:analytic:reordered}
\end{align}

Solving for $\DD$ and $\vS_2$, the explicit fluid-fluid reflection matrix is obtained as:

\begin{equation}
	[\mbf{R}] =
	\begin{bmatrix}
		-1/Z_2 & 2 &0\\
		1/Z_2& -1 &0
	\end{bmatrix}
	\label{eq:ex:R}
\end{equation}

In this matrix, the last null column corresponds to the null-valued characteristics which
are, by definition,  non-propagative and are not transmitted through the coupling
interface. The coefficients of $[\mbf{R}]$ are generally difficult to analyze (since they
couple two very different approaches). The consistency of the form derived may be checked
through the diagonal terms. For a fluid-fluid coupling, the first diagonal term
corresponds to the ratio of acoustic velocity and pressure and fits with the definition of
acoustic impedance: $Z = p/v$.  Finally, the second diagonal term makes sense in the DGM
framework since it forces the incoming and outgoing characteristics to propagate in
opposite directions.

	\subsection{Remarks on discretization and choice of shape functions}
	\label{sec:coupling:impl}

Equation~\eqref{eq:dgm:adjoint} sets a requirement on the set of trial functions without
actually proposing an expression for them. In the present work, following
~\cite{gabard_discontinuous_2007}, the test functions are chosen to be in the form of
plane waves.  Thus, they are composed as a superposition of $N_w$ plane waves whose
directions are distributed in the unit disc . Using this interpolation, the $\vS$ vector
on each element is expanded as:

\begin{equation}
	\vS = \sum_{n=1}^{N_w} a_n \mbf{U}_ne^{-j\mbf{k}_n\cdot(\mbf{r}-\mbf{r}_c)}
	\label{eq:dgm:pwbasis}
\end{equation}
with $\mbf{r}_c$ mapping to the center of the element and $\mbf{k}_n$ and are the wave
numbers of the different plane waves. The $\mbf{U}_n$ vector acts in a fashion similar to
that of polarization vectors, relating the propagating term to certain quantities of the
state vector.  Finally, the set of the different $a_n$ for all the elements are the
amplitude factors and components of the solution vector. Using such an interpolation
strategy for the state vector, the method is called DGM with plane waves (PWDGM). A more
in-depth explanation is given in reference~\cite{gabard_discontinuous_2007}.

\begin{figure} 
\centering
	\includegraphics{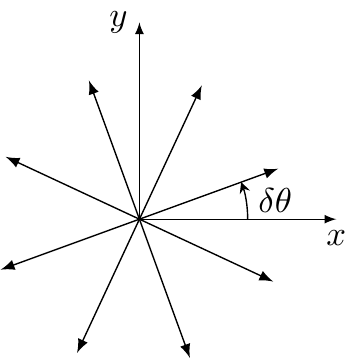}
	\caption{Example of recontruction basis for PWDGM with $N_w=8$. The tilt of the basis
	with respect to the $x$ axis is controled by $\delta\theta$}
	\label{fig:pwbasis}
\end{figure}

As a finite number $N_w$ of plane waves is distributed on the unit disc, as shown in
figure~\ref{fig:pwbasis}, one can easily understand it induces privileged directions. Each
time the solution's direction of propagation coincides with one of the waves of the basis,
one expects the error to drop. The reconstruction of waves whose directions are situated
between two waves of the basis is challenging. A way to ensure a good approximation for
all directions is to use a $N_w$ sufficiently large, reducing the distance between two
waves and introducing more directions where error drops (privileged). A second possibility
is to tilt the wave basis using $\delta\theta$ and align one of the wave with the
solution's direction of propagation. This can be done independently from the other
elements' basis but finding a good algorithm to predict the best tilt is not a trivial
task. Another way to address this precision loss is to use more elements, disposed so to
avoid patterns in the mesh and take advantage of the inter-element
compensation~\cite{gabard_discontinuous_2015}.  The main limit to this approach then comes
from the lower bound in element size: as the solution is reconstructed by superposing
plane waves, they must be given enough space between their origin (center of element) and
the boundaries to combine.

On the other hand, the usual way to interpolate fields in the scope of FEM makes use of
polynomials. The edges of the mesh are augmented of $p+1$ nodes (including both ends) used
for a $p$-order approximation. Different basis have been proposed over the years. For the
examples that follow, quadratic Lagrange polynomials were used.

In the light of the previous considerations, it is clear that all the integral terms
from~\eqref{eq:coupling:final} require numerical integration of polynomials and/or
exponentials. When the integrals only involve exponentials or polynomials, as for the
first and last terms, one can use a fully analytical	integration. Even if possibly
cumbersome to write, the numerical evaluation of the integral will then be as precise as
the machine itself and computationally efficient.

Nevertheless, when considering integrals involving products of polynomials and
exponentials (as do the second and third terms of~\eqref{eq:coupling:final}), the
integration scheme's performance is critical. Proposing an efficient numerical
integration of products of polynomials and exponentials is not trivial. Complex
exponentials and polynomials tends to oscillate at different scales, rendering difficult
the choice of a integration technique. Finally the large number of integration points
required to get a proper value tends to slow down the resolution: indeed, the integration
is triggered twice per coupling segment. Some improvements in the medium-high frequency
range may be accessible through a change of integration technique or pre-computing.

	\section{Application and results}
	\label{sec:testing}

To validate the proposed coupling approach as well as to demonstrate its performance, some test cases are presented in this section. The first two provide
insights into the convergence and dispersion properties and an existing analytical solution is used as a reference.
The quality of the results is evaluated through $\mathcal{L}_2$ relative error:

\begin{equation}
	\epsilon = \bigg(\frac{
		\int_\Omega \left|p-p_{ref}\right|^2\dd\Omega
	}{
		\int_\Omega \left|p_{ref}\right|^2\dd\Omega
	}\bigg)^{1/2}
	\label{eq:testing:l2error}
\end{equation}
where $p$ corresponds to the predicted pressure and $p_{ref}$ to the analytical reference.
The third example deals with a more complex geometry and aims at demonstrating the hybrid method with
incompatible meshes, acute corners, rounded shape and a relatively large domain.
In all simulations air medium at 20\textdegree C is used with the following properties:

\begin{equation}
	\rho = 1.213~\mathrm{kg}\cdot\mathrm{m}^{-1}, \qquad c = 341.973~\mathrm{m}\cdot\mathrm{s}^{-1}
\end{equation}

In the examples, $\delta\theta$ is always null except otherwise stated. 

	\subsection{Dispersion analysis}
	\label{sec:testing:disp}

\begin{figure} 
\centering
	\includegraphics{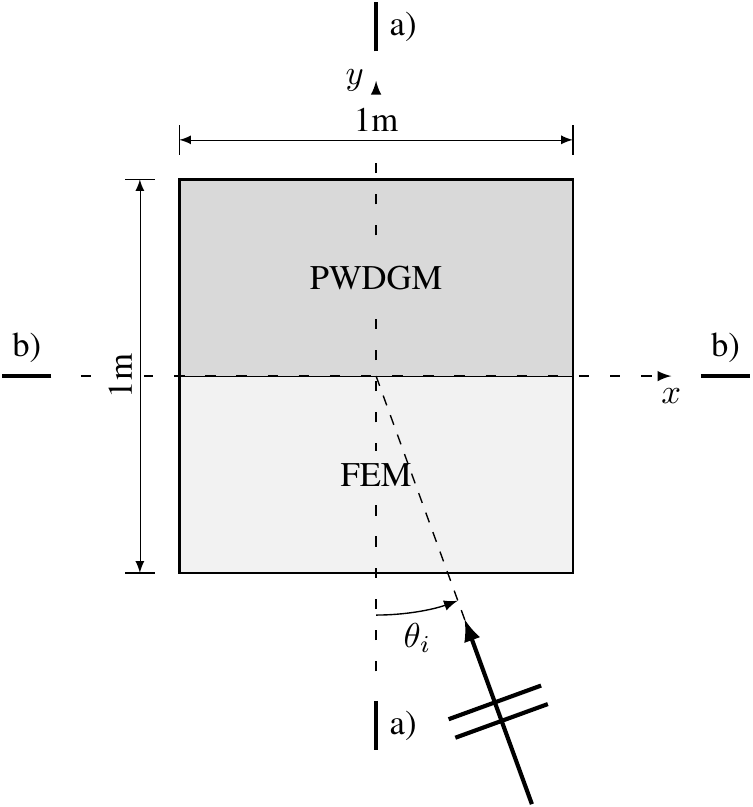}
	\caption{The domain used for the dispersion analysis, a $1$m square with the plane
	wave field imposed on all sides at an angle $\theta_i$. Positions a) and b) are of
	special interest.}
	\label{fig:dispersion}
\end{figure}

For the dispersion analysis a square domain with 1\,m side length is used, see
figure~\ref{fig:dispersion}. The purpose is to demonstrate the numerical behaviour of the
proposed coupling approach. The excitation is in the form of a plane wave imposed on the
boundary of the square, the wave front propagating at an angle $\theta_i$ from the lower
$y$ axis.

In figure~\ref{fig:dispersion}, two different angles of particular interest are identified
as a) and b). The former corresponds to a wave entering through the boundary of one of the
sub-domains only, implying  that the error builds up sequentially in one sub-domain first
and then the other. For directions marked b) the incident field is parallel to the
coupling boundary and there is no flux across the interface for the reference solution.

A pure PWDGM solution is also computed with the purpose of providing a numerical result
which is independent of the proposed coupling approach. This solution is computed for the
same mesh and approximation order as used in the PWDGM part of the coupled model.

\begin{figure} 
\centering
	\includegraphics[width=.8\textwidth]{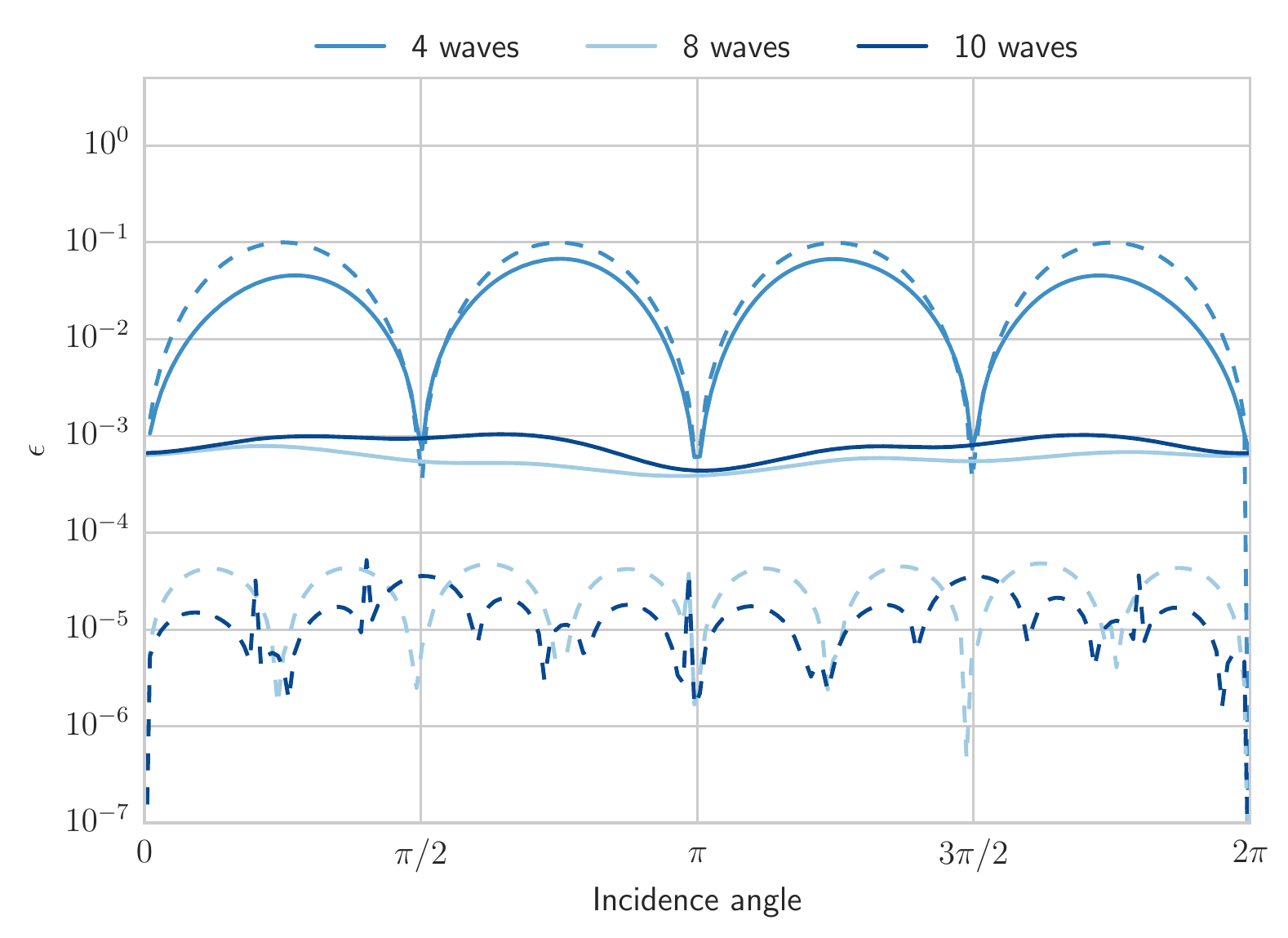}
	\caption{Evolution of the relative error with respect to the incidence angle of a
	plane wave over a 1m square domain. Half the domain is modelled with DGM and the rest
	with FEM. Triangular elements are used on both sides, dashed lines are for the pure DGM
	reference and $f=1000$\,Hz. (Color online)}
	\label{fig:disp:tri}
\end{figure}

From a theoretical point of view, the FEM part of the solution is not sensitive to the
angle of incidence and the error as a function of $\theta_i$ should be constant.  In
practice, the mesh may have an influence, introducing privileged directions where the
error may build up but this is generally negligible. The solution obtained using only PWDGM for the square problem
exhibits $N_w$ arches over the $[0,~2\pi]$ range of incidence angle. This solution is
referred to as the pure PWDGM reference in figure~\ref{fig:disp:tri}. The observed arches
are well known in the literature, and stem from the privileged directions induced by the
waves used in the base (see section~\ref{sec:coupling:impl} and figure~\ref{fig:pwbasis}).
However, as the number of waves is increased the error related to these is reduced, see
figure~\ref{fig:disp:tri}.

In figure~\ref{fig:disp:tri}, the results for the coupled FEM-DGM solution at $f=1000$\,Hz
are shown. With 4 waves per element, the archs discussed above appear in the solution, but
are not observed for the higher number of waves, $N_w=8$ and $N_w=10$.  The error
seems to be controlled by the FEM solution, as the overall accuracy for this coupled
solution is about two orders of magnitude worse than the pure PWDGM solution. Furthermore,
the error levels for $N_w=8$ and $N_w=10$  and above are close to each other, indicating
that the FEM is limiting the accuracy in this case and this observation is confirmed by a
direction-independent error around $10^{-3}$.

\begin{figure} 
\centering
	\includegraphics[width=.8\textwidth]{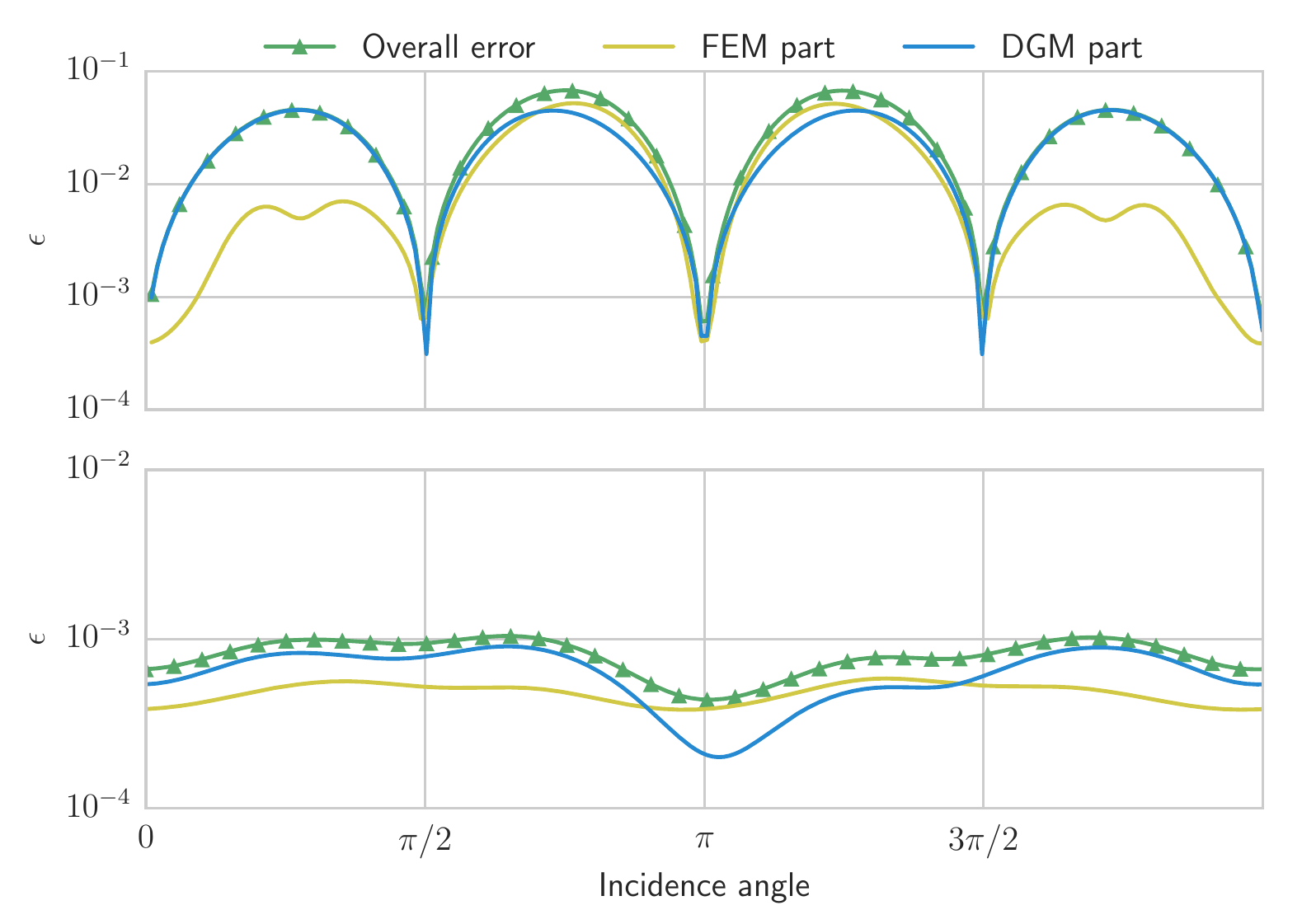}
	\caption{Contribution of the two different schemes to the global error  with respect
	to the incidence angles for $f=1000$\,Hz and $N_w=4$ (top) or $N_w=10$ (bottom). (Color Online)}
	\label{fig:disp:contribs}
\end{figure}

To investigate this further the error
generated in the FEM and PWDGM domains separately is shown in figure~\ref{fig:disp:contribs} for $Nw \in \{4, 10\}$.

For $N_w=4$, the PWDGM-related error is dominating over the whole range. Between
$\theta_i=\pi/2$\,rad and $\theta_i=3\pi/2$\,rad, where the error is generated first in the PWDGM
domain before being transmitted to the FEM and both methods then exhibit the same error profile. When comparing with the
error in the FEM domain for the other half of the range (wave coming from the FEM side),
one can see that the PWDGM archs are actually preponderant in shaping the error profile.
For this first case, the hybrid method has an error level similar to that of the pure PWDGM reference.

For $N_w = 10$, while figure~\ref{fig:disp:tri} shows  that PWDGM is not limiting
the accuracy, the different contributions at the bottom of figure~\ref{fig:disp:contribs},
suggests that both methods contribute more or less with equal shares in the error
generation.

This apparent contradiction is related to the large difference of precision between a
10 wave PWDGM and FEM for the coupled system. As figure~\ref{fig:disp:tri} suggests that
the PWDGM should be accurate enough and the error thus has its origin in the FEM part with
its lower resolution in the model used. The error is dominated by phase error
building in the FEM which in the coupling to the PWDGM  pollutes the PWDGM solution for these
particular directions of incidence.

	\subsection{Kundt's tube}
	\label{sec:examples:kundtconv}

\begin{figure} 
	\centering
	\includegraphics{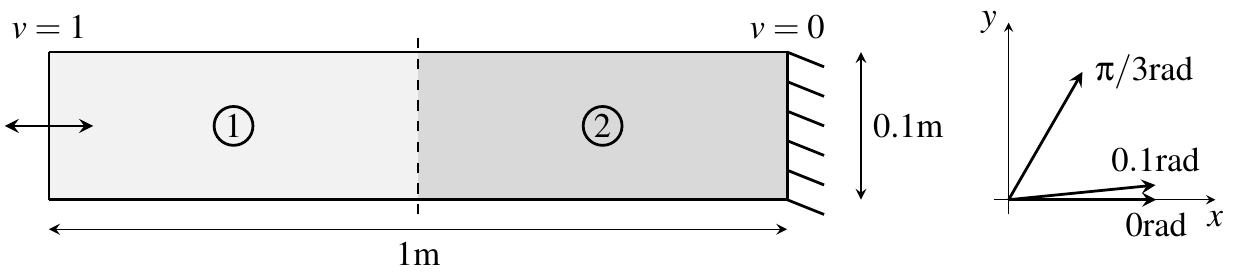}
	\caption{A simple 1$\times$0.1m Kundt tube used as a benchmark. The two labeled
	regions are modelled with FEM or PWDGM. On the right, the first wave of the basis used
	for figures~\ref{fig:kundt:convergence} to~\ref{fig:kundt:dispnx}.}
	\label{fig:kundt}
\end{figure}

In this second example, the proposed coupling  method is used to predict the pressure
field in a rigid, rectangular cavity excited by a unit velocity at one end, see
figure~\ref{fig:kundt}. This cavity has dimensions 1\,m$\times$0.1\,m and is divided into two
sub-domains with a interface at the middle of the tube. The purpose is to assess the
DGM sensitivity to the wave base, here through varying the tilt angle of the
base as shown in figures~\ref{fig:pwbasis} and~\ref{fig:kundt}.

\begin{figure} 
	\centering
	\includegraphics[width=.9\textwidth]{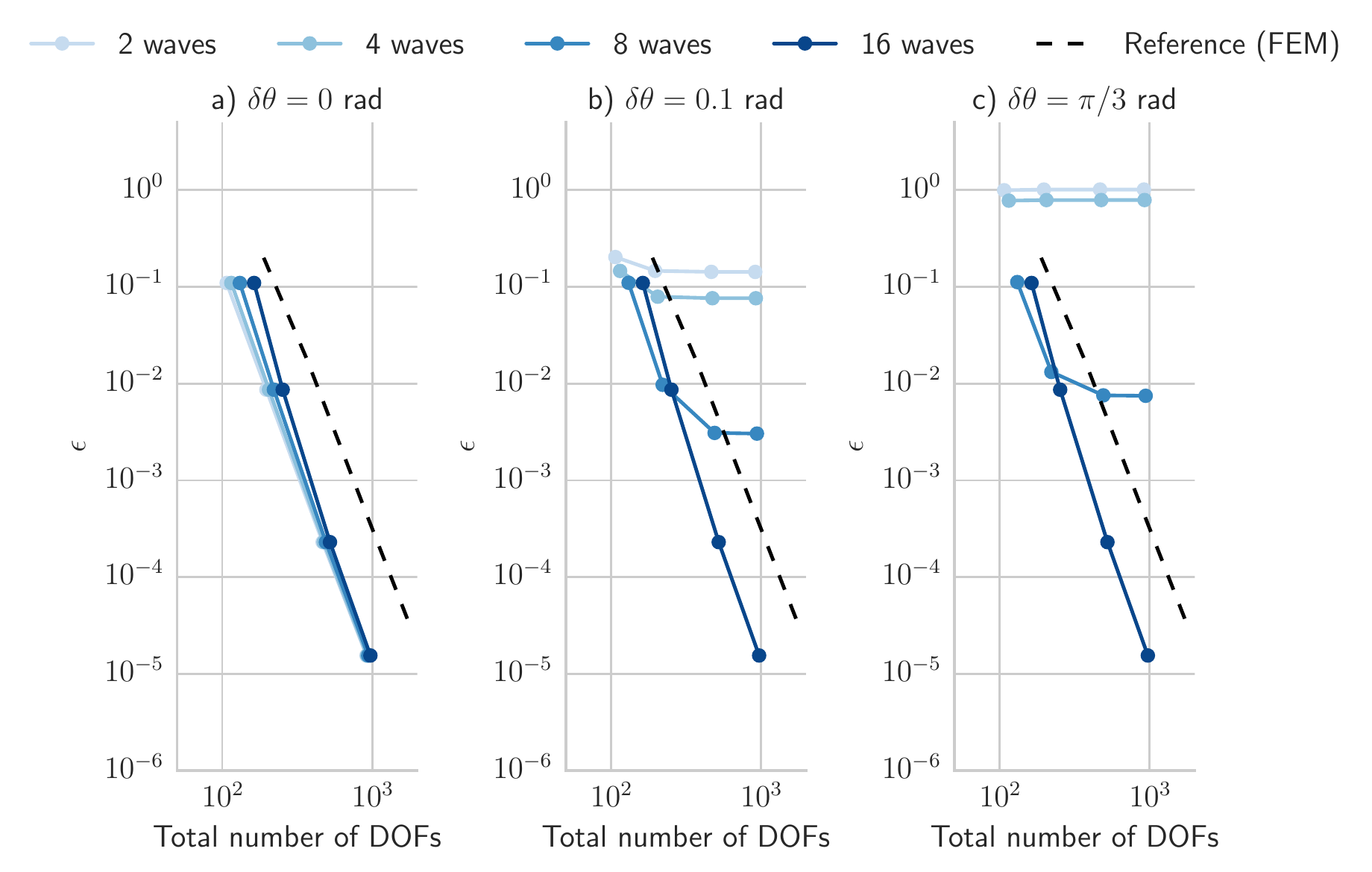}
	\caption{Evolution of the relative $\mathcal{L}_2$ error with respect to the the size
	of the linear system for $f=1000$\,Hz. The dashed line is a reference for which the all
	domain was modelled using FEM ; other lines are for different $N_w$. (Color online)}
	\label{fig:kundt:convergence}
\end{figure}

To assess the hybrid model's convergence rate, a baseline case with the whole domain
modelled with FEM is used. This is shown as the black dashed line in figure
~\ref{fig:kundt:convergence}. In the numerical tests performed with the hybrid model one
of the sub-domains is modelled by FEM and the other by PWDGM. Initial checks with
excitation applied at both ends did not reveal any noticeable difference, thus only
results for an excitation applied on the FEM side are included here.

\medskip

This two first tests aim at testing the two classical refinement approaches for FEM and
PWDGM on the hybrid model. The first of the tests uses a classical FEM strategy by
gradually refining the FE side mesh without modifying the PWDGM parameters. The analysis
conducted for different numbers of waves $N_w$ in the reconstruction basis and different
tilt angles $\delta\theta$. The refinement of the FE sub-domain is the same as the on used
to compute the baseline (dashed black line) shown in figure~\ref{fig:kundt:convergence}.
Each of the three graphs is computed for a noticeable tilt $\delta\theta$ of the wave
basis. Indeed, for a long, merely 1D, acoustic cavity, keeping one wave aligned with the
main axis of the cavity seems beneficial. In order to understand this behaviour, the
properties of the interpolation strategy for the PWDGM as explained in
section~\ref{sec:coupling:impl} are useful. For example, figure
\ref{fig:kundt:convergence}a is obtained for a highly favourable PWDGM configuration.
Since an even number of waves $N_w$ is considered, $N_w/2$ pairs corresponding to the two
propagation directions of the solution (forward \& backward) are present and one of them
aligned with the solution direction.  This allows to decrease considerably the
reconstruction error in the PWDGM domain, down to machine precision. In this case, no
error should be generated by the PWDGM and only the FEM sub-domain and coupling have an
impact.  In figure~\ref{fig:kundt:convergence}.a), the convergence rate of the hybrid
method for all numbers of waves in the basis is comparable to that of the pure FE baseline
(black dashed line). This first result indicates that the coupling itself is not
negatively contributing to the error.

\smallskip

For the two last parts of figure~\ref{fig:kundt:convergence}, a saturation of the error
rate can be observed, even on figure~\ref{fig:kundt:convergence}b where the tilt angle is
small. This behaviour complies with the PWDGM effect previously discusssed.  When the FE
part of the hybrid scheme reaches at a sufficently high refinement factor, the PWDGM error
can become the highest and the overall error doesn't drop any further when adding elements
in the FEM domain.  The main difference between graphs~\ref{fig:kundt:convergence}b and
\ref{fig:kundt:convergence}c is to be seen for $N_w\in\{2,4\}$ compared to $N_w=8$. Indeed
for small a number of waves tilting the basis has a dramatic impact (resulting in 100\%
error rate) whereas the other waves compensate the precision loss when there are
sufficiently many in the base ($N_w=8$ here).

Finally, the error for $N_w=16$ stays the same for all tilt angles $\delta\theta$. An
explanation for it is to be found in the relative error levels for the FEM and PWDGM
domains. Adding waves to the basis tends to reduce the maximum error in PWDGM
~\cite{gabard_discontinuous_2007,gabard_discontinuous_2015} and, once a given $N_w$ is
reached, the PWDGM error does not exceed the FEM error anymore. For $N_w=16$, the PWDGM
error is systematically lower than the FEM error which then dominates regardless of the
orientation of the solution with the PWDGM reconstruction basis. As the FEM solution is
not sensitive to the tilt of the wave basis, the convergence curve is not impacted either
and the hybrid scheme converges at the same rate as the FEM for all $\delta\theta$.

\bigskip

\begin{figure} 
\centering
	\includegraphics[width=.8\textwidth]{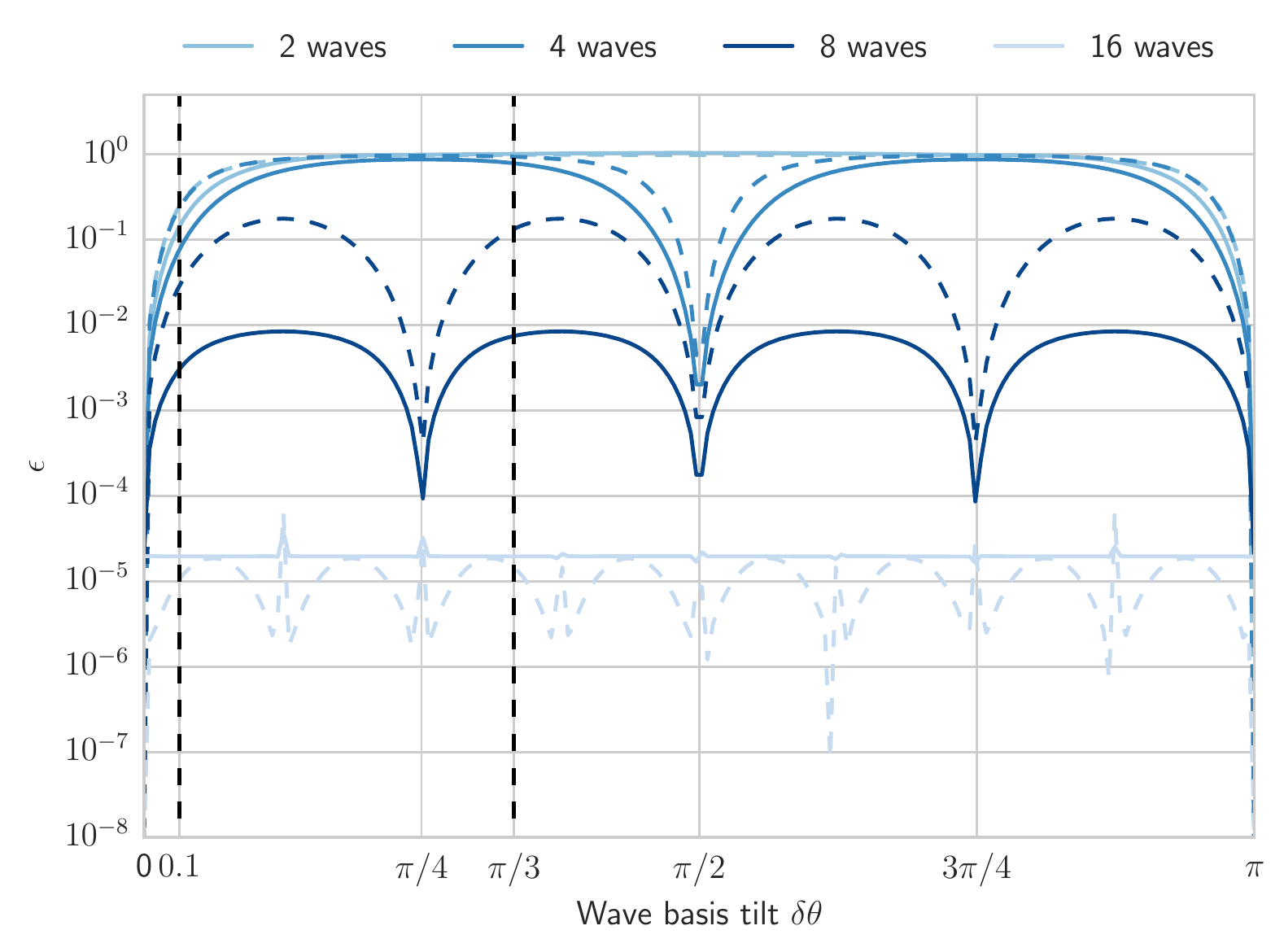}
	\caption{Evolution of the relative error with respect to the tilt angle
	$\delta\theta$ for different numbers of waves and a given
	refinement of the FE mesh at $f=1000$\,Hz. Dashed lines are the corresponding full DGM
	reference. The two vertical dashed lines mark the two angles used on
	figure~\ref{fig:kundt:convergence}. (Color online)}
	\label{fig:kundt:dispnw}
\end{figure}

Keeping the number of waves $N_w$ constant and varying the FE mesh, and vice versa,
the influence of the tilt angle $\delta\theta$ on the error in the hybrid solution may be
studied. For a fixed FE mesh, and varying of the number of waves in the base, the results
shown in figure~\ref{fig:kundt:dispnw} where $\delta\theta$ varies between $0$ and $\pi$,
are obtained. The tilt angles used in the previous results,
$\delta\theta=0.1$\,rad and $\delta\theta=\pi/3$\,rad, are marked in the
figure. Results for different $N_w$ are presented in figure~\ref{fig:kundt:dispnw} with 2
superimposed black dashed lines for $\delta\theta=0.1$\,rad and
$\delta\theta=\pi/3$\,rad.  The results from the hybrid method (solid lines) are
compared with their pure PWDGM counterparts (dashed lines with matching color). The results
confirm the previous observations in the current paper and are consistent with previously
published research, ~\cite{gabard_discontinuous_2015}. The PWDGM error is at its lowest
whenever a wave of the basis is aligned with the solution. This is observed for both the
pure PWDGM baseline references and the coupled solutions. For the latter though, the peak
error is lower as a consequence of the error build-up mechanisms in a PWDGM domain.

\begin{figure} 
\centering
	\includegraphics[width=.8\textwidth]{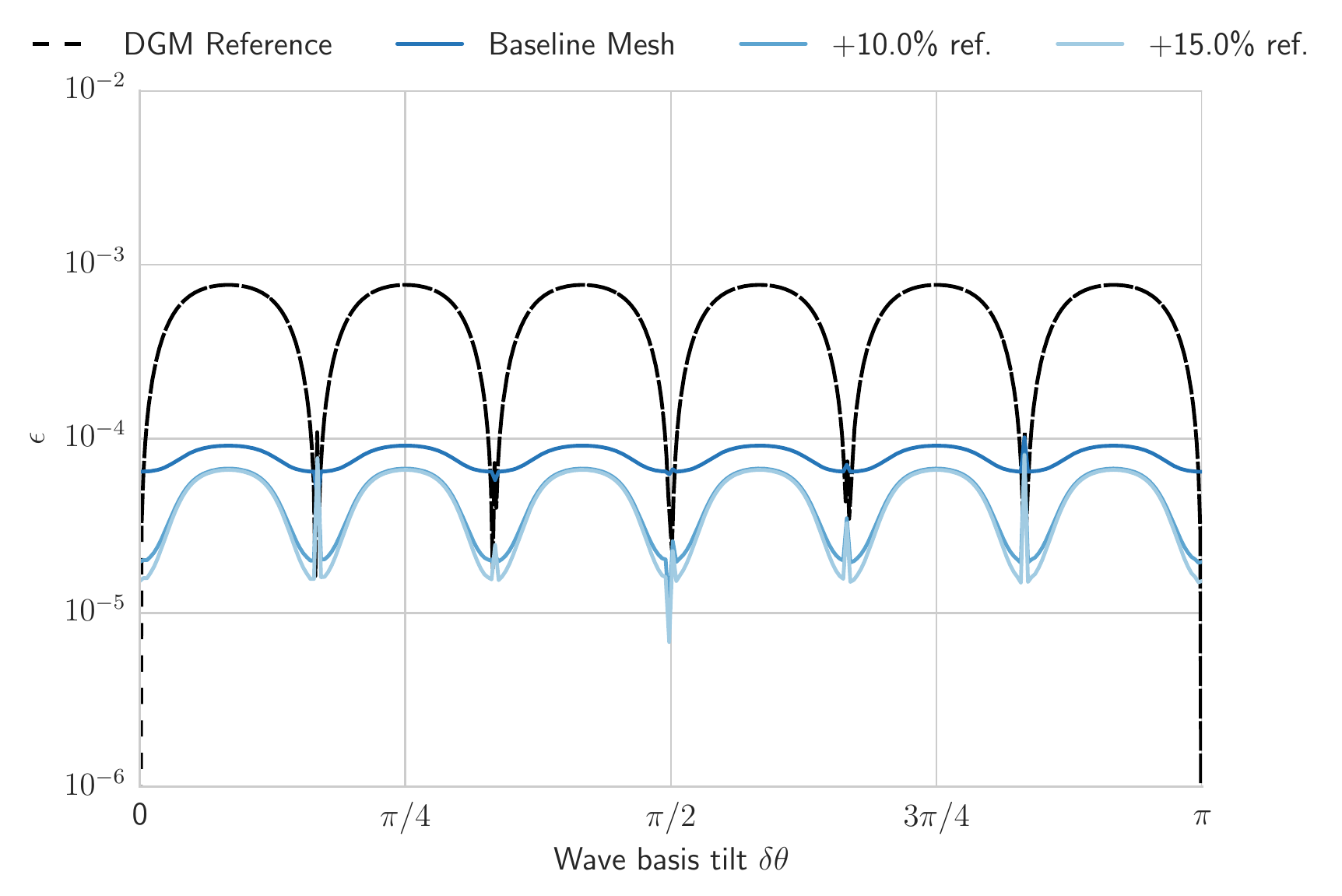}
	\caption{Evolution of the relative error with respect to the tilt angle
	$\delta\theta$ for a given number of waves but different refinements of the FE mesh
	(the lighter the more refined). The graph was generated at $f=1000$\,Hz and the black
	line is a pure DGM reference. (Color online)}
	\label{fig:kundt:dispnx}
\end{figure}

For a fixed number of waves $N_w$, the influence of the mesh refinement is shown in
figure~\ref{fig:kundt:dispnx}. Three different refinements of the FE mesh (baseline and 10
or 15\% increase in number of degrees of freedom) are shown as function of $\delta\theta$.
This graph shows that replacing half of the domain by a FE model allowed to reduce the
maximum error. On the other end, it also suggests than if the added FEM is not precise
enough, it tends to prevent the error drops introduced by PWDGM.

To add some background to this phenomenon, one could go back to the PWDGM solution as such.
The PWDGM establishes a flux through the boundaries of the elements and the associated
errors are related to the quality of the reconstruction of the flux at these boundaries.
If half the domain is replaced by another (potentially more precise) method, then part of
the DG error cannot build up and the global error ends up capped to a lower rate. For
$N_w=8$ in figure~\ref{fig:kundt:dispnw}, the effect is particularly prominent, a full PWDGM solution gives a $10^{-1}$ peak error but coupling to FE allows to drop down by one order of magnitude.

	\subsection{Open resonant cavity with tight corner}

\begin{figure} 
	\centering
	\includegraphics{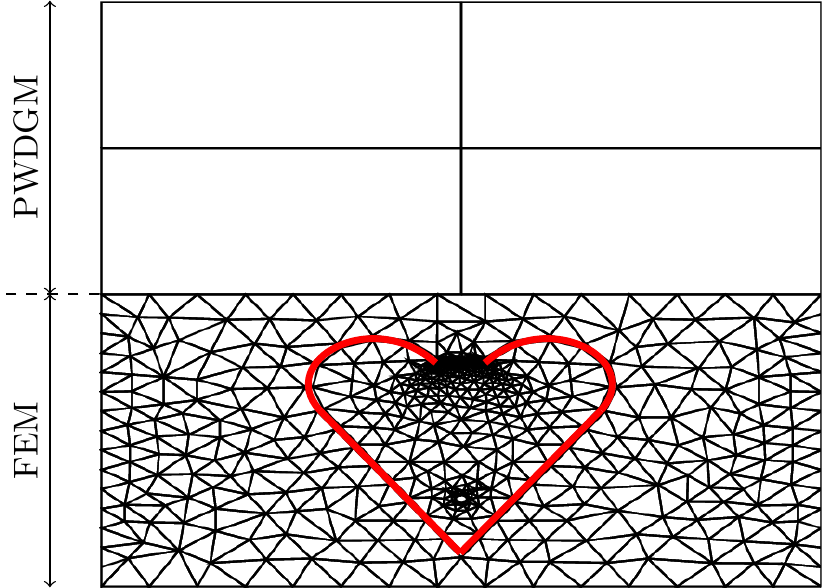}
	\caption{Example of mixed mesh with an incompatible coupling interface.}
	\label{fig:mesh_example}
\end{figure}

To demonstrate the applicability of the proposed coupling approach, a hybrid solution
of a slightly more complex geometrical arrangement is used. A 1 meter-square cavity is
filled with air and an open resonant rigid body is placed inside. The lower part,
containing the resonator, is modelled using FEM in order to accurately account for all the
geometrical details. As an example, the sharp angle at the lowest part of the rigid body
makes the field inside hard to resolve using plane waves. In addition, the top part is
modelled using 4 PWDGM elements, see figure~\ref{fig:mesh_example}, with a base of 32 plane
waves per element.

\begin{figure} 
	\hspace{-0.07\textwidth}%
	\begin{subfigure}[b]{.61\textwidth}
		\includegraphics[width=\textwidth]{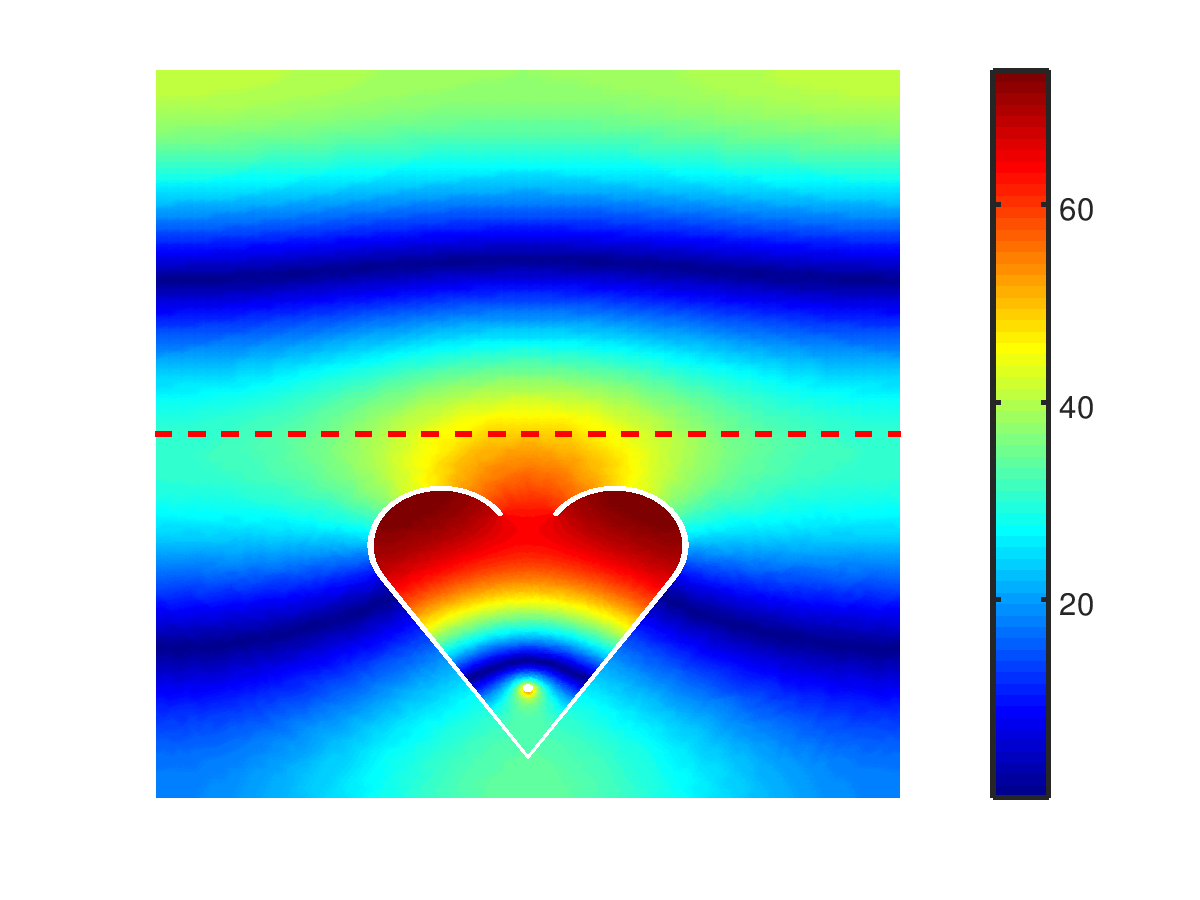}
			\caption{Excitation by a point source placed\\near the bottom corner of the
			resonator with $f=52$\,Hz.}
		\label{fig:heart:result_ps}
	\end{subfigure}\hspace{-0.03\textwidth}%
	\begin{subfigure}[b]{0.6\textwidth}
		\includegraphics[width=\textwidth]{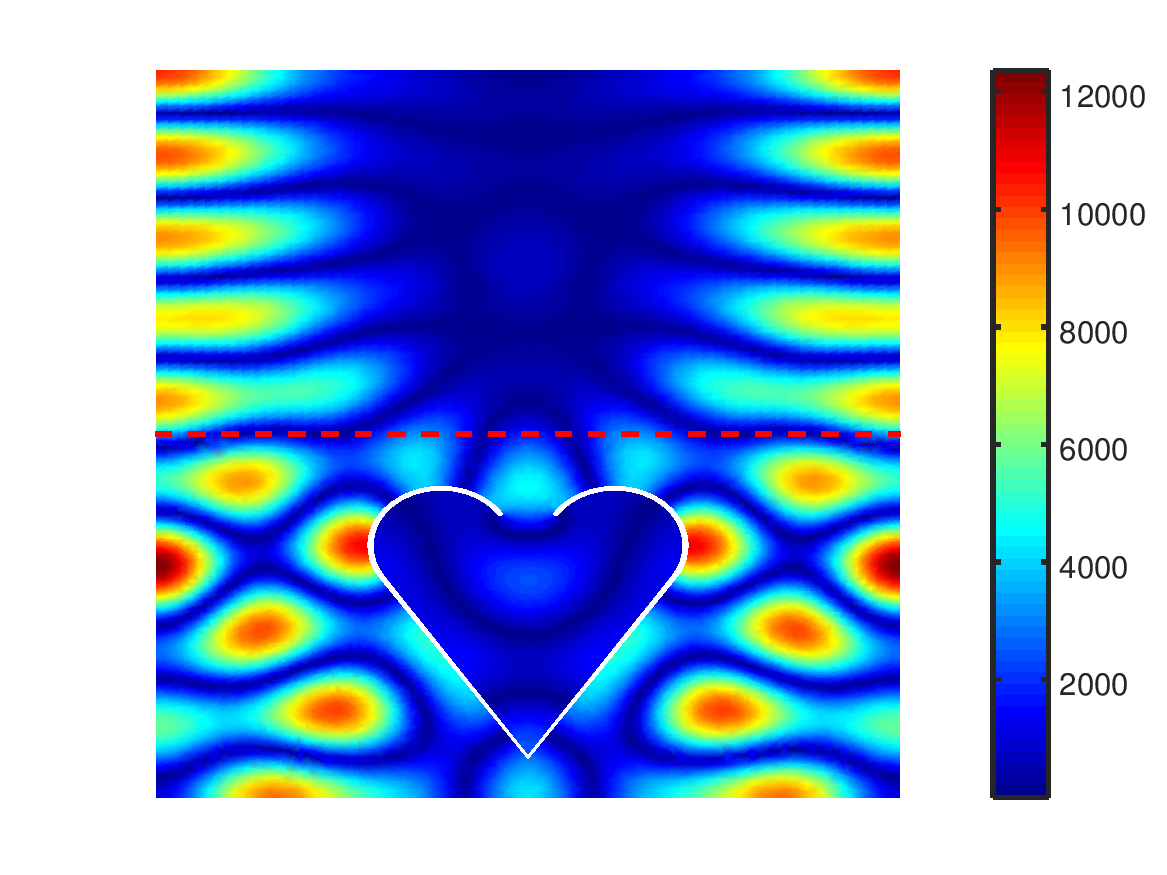}
			\caption{Excitation by a unit velocity from the top,\\ rigid boundary with
			$f=260$\,Hz.}
		\label{fig:heart:result_te}
	\end{subfigure}
	\caption{Simulation of the pressure field for two different excitations. Compared to
	the one presented as an example in figure~\ref{fig:mesh_example}, only the FEM
	domain's refinement is increased. The dashed line symbolizes the coupling interface.}
\end{figure}

Two different cases are tested. First, a point source is placed at the bottom of the
resonator and excites the volume through a unit velocity at $f=52$\,Hz. The
resulting pressure field for this case is shown in figure~\ref{fig:heart:result_ps}.
The second test case uses an unit velocity excitation from the top of the cavity, at
$f=260$\,Hz. The pressure map for the second example is shown in
figure~\ref{fig:heart:result_te}. In both figures, the dashed red line marks the
interface $\Gamma$ between the two sub-domains, each modelled using the two different
discretization methods, coupled using the approach discussed in the present paper.
For both studied case, the naive replacement of half the FEM domain by a PWDGM one led to
a 15\%  decrease of the number of degrees of freedom in the final linear system.

\begin{figure} 
	\centering
	\includegraphics[width=\textwidth]{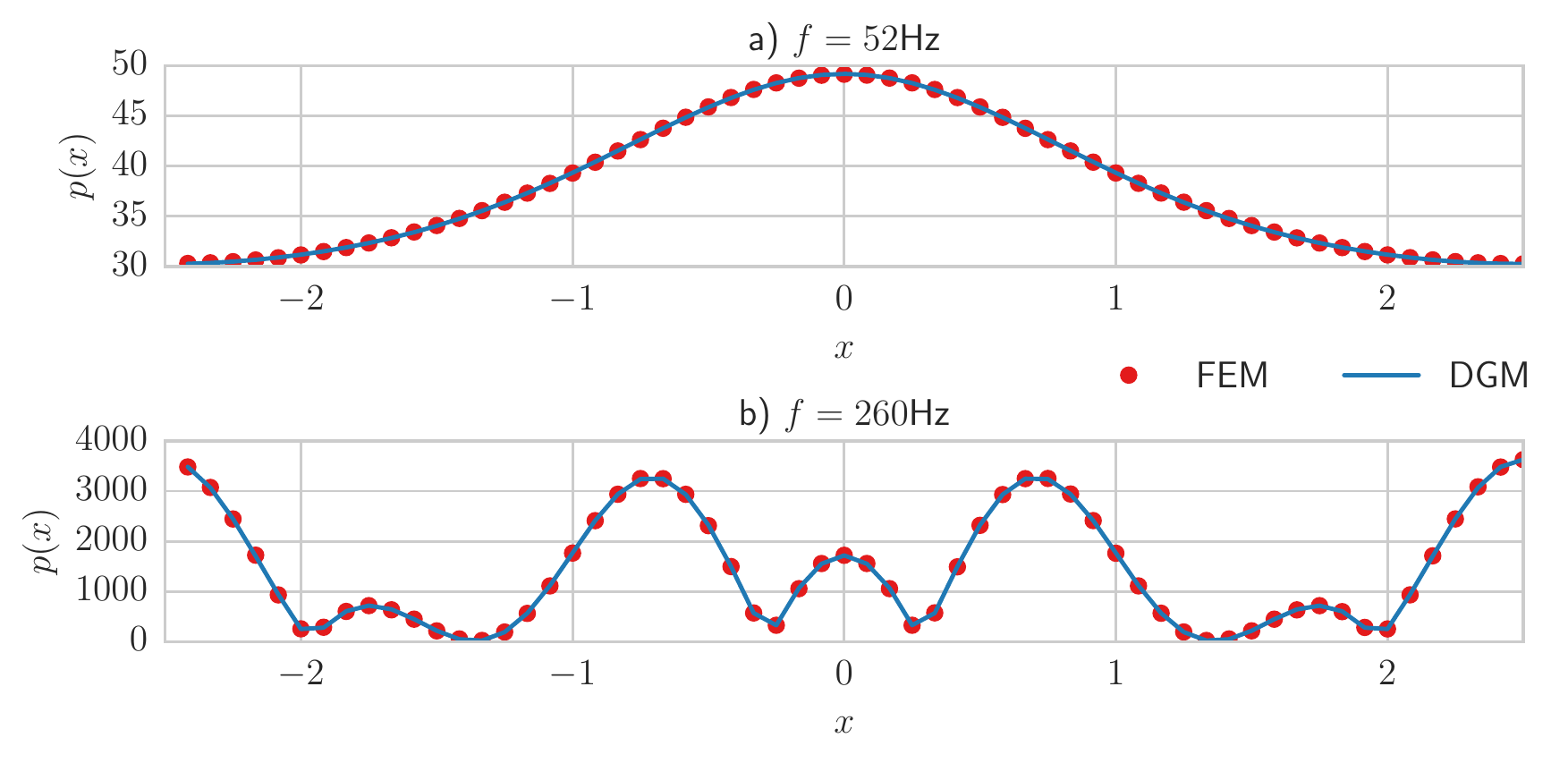}
	\caption{Pressure profiles along the interface between the method as seen by the
	FE and PWDG schemes. (Color online)}
	\label{fig:heart:jump}
\end{figure}

As may be seen from figure~\ref{fig:heart:result_te}, the pressure solutions is visually
continuous with no artefacts generated close to the coupling interface. Thus, the pressure
field, generated in the FE domain seems correctly transmitted to the PWDGM domain and the
cavity modes seem to be reconstructed correctly in both sub-domains despite the coarse
PWDGM mesh. As part of the investigations made, a converged pure FEM resolution
produced the same results. Interestingly, for the higher frequency excitation, the coarse
DGM mesh captures the lobes well.

\medskip

In order to demonstrate that the continuity between the pressure at the interface between
the sub-domains is fulfilled, the pressure along $\Gamma$ as computed by each of the two
methods is shown in figure~\ref{fig:heart:jump}. For the first case, one clearly sees that
the central lobe, due the resonator's radiation, is approximated the same way from both
sides. For the second one, the same pressure profile is computed even if many modes are
spread across the boundary. These results confirm the validity of the proposed coupling
approach.


	\section{Conclusion}

This work proposes an effective coupling strategy for the Finite-Element Method and the
Discontinuous Galerkin with Plane Waves. The resulting hybrid scheme is based on a
reflection matrix that maps FEM's physical field and DGM's characteristics seamlessly.
This procedure presents several advantages, the first of which being to enforce a
discrepancy-free coupling. The coupling then relies on continuity conditions between state
vectors that are
straightforward to write for most pairs of media. The litterature on FEM adaptations of
physical models is abundant (even for complex media as in~\cite{atalla_enhanced_2001}) and
extensions the PWDGM approach has been proposed~\cite{gabard_discontinuous_2015}
recently.

After presenting the coupling and an analytical derivation for a simple case, numerical
examples were provided. First an set of academic examples were used to support the claim
of a incidence angle-independant coupling, methods specificities conservation and
simulation ability. Both methods limits were explored in the scope of the coupled approach
in these first examples and it was made clear how this approach could be used optimally in
a third one.

The last use case showed how FEM could be used to protect resonant features or geometrical
details and embed them in a larger PWDGM domain. The good adaptativity of the first method
was combined to the large-scale modelling properties of the second to reduce model
complexity without precision loss.


	\bibliography{FEMDGM}{}
	\bibliographystyle{wileyj}

  %

	 \appendix
  %

	\newpage

\end{document}